%% file: main.tex
\documentclass[sigconf]{acmart}

\usepackage[utf8]{inputenc}
\usepackage[T1]{fontenc}
\usepackage{microtype}
\usepackage[a-2b]{pdfx}
\usepackage{graphicx}
\usepackage{balance}  %
\newcommand\norm[1]{\left\lVert#1\right\rVert}
\usepackage{subfigure}
\usepackage{multirow}
\usepackage{algorithm}
\usepackage{algorithmic}
\usepackage{color}
\usepackage{listings}
\usepackage{amsmath}
\usepackage{float}

\usepackage{amsmath}
\usepackage{subscript}
\usepackage[skip=0pt]{caption}
\usepackage[labelfont=bf]{caption}
\newcommand{\dif}[1]{\mathop{}\!\mathrm{d}#1\mathop{}}

\newcommand{\eat}[1]{}

\AtBeginDocument{%
  \providecommand\BibTeX{{%
    \normalfont B\kern-0.5em{\scshape i\kern-0.25em b}\kern-0.8em\TeX}}}

\setcopyright{acmcopyright}
\copyrightyear{2018}
\acmYear{2018}
\acmDOI{10.1145/1122445.1122456}

\acmConference[Woodstock '18]{Woodstock '18: ACM Symposium on Neural
  Gaze Detection}{June 03--05, 2018}{Woodstock, NY}
\acmBooktitle{Woodstock '18: ACM Symposium on Neural Gaze Detection,
  June 03--05, 2018, Woodstock, NY}
\acmPrice{15.00}
\acmISBN{978-1-4503-XXXX-X/18/06}

\begin{document}

\title{Benchmark of DNN Model Search at Deployment Time}

\author{Lixi Zhou}
\affiliation{%
  \institution{Arizona State University}
  \city{Tempe}
  \country{USA}}
\email{lixi.zhou@asu.edu}

\author{Arindam Jain}
\affiliation{%
  \institution{Arizona State University}
  \city{Tempe}
  \country{USA}}
\email{ajain243@asu.edu}

\author{Zijie Wang}
\affiliation{%
  \institution{Arizona State University}
  \city{Tempe}
  \country{USA}}
\email{zijiewang@asu.edu}

\author{Amitabh Das}
\affiliation{%
  \institution{Arizona State University}
  \city{Tempe}
  \country{USA}}
\email{adas59@asu.edu}

\author{Yingzhen Yang}
\affiliation{%
  \institution{Arizona State University}
  \city{Tempe}
  \country{USA}}
\email{yyang409@asu.edu}

\author{Jia Zou}
\affiliation{%
  \institution{Arizona State University}
  \city{Tempe}
  \country{USA}}
\email{jia.zou@asu.edu}

\begin{abstract}
Deep learning has become the most popular direction in machine learning and artificial intelligence. However, the preparation of training data, as well as model training, are often time-consuming and become the bottleneck of the end-to-end machine learning lifecycle.
Reusing models for inferring a dataset can avoid the costs of retraining. However, when there are multiple candidate models, it is challenging to discover the right model for reuse. Although there exist a number of model sharing platforms such as ModelDB, TensorFlow Hub, PyTorch Hub, and DLHub, most of these systems require model uploaders to manually specify the details of each model and model downloaders to screen keyword search results for selecting a model. We are lacking a highly productive model search tool that selects models for deployment without the need for any manual inspection and/or labeled data from the target domain. This paper proposes multiple model search strategies including various similarity-based approaches and non-similarity-based approaches. We design, implement and evaluate these approaches on multiple model inference scenarios, including activity recognition, image recognition, text classification, natural language processing, and entity matching.
The experimental evaluation showed that our proposed asymmetric similarity-based measurement, adaptivity, outperformed symmetric similarity-based measurements and non-similarity-based measurements in most of the workloads.
\eat{We identified a technical challenge that the matching relationship between a target query dataset and a training dataset is asymmetric. Therefore, we proposed a new adaptivity metric, and a novel two-level locality sensitive hashing index to facilitate the comparison of the adaptivity of the target query dataset to a number of models' training datasets in order to search for the proper model to reuse.}
\end{abstract}



\maketitle

\input{intro}

\input{insights}
\input{background}

\input{solution}
\input{evaluation}

\input{relatedworks}

\section{Conclusions}
In a broad class of AI applications, such as Uber's marketplace-level forecasting and downtime prediction for manufacturers, a large number of model versions will be created for different locations and different time points. To facilitate model reuse,  we systematically explore the problem of searching models from a repository of pre-trained models that have similar architectures, for serving on a target query dataset. We are particularly interested in solving the problem for fast model deployment scenarios, where little labeled data is available. To address the problem,  we propose five techniques, including symmetric similarity-based approaches, an asymmetric similarity-based approach utilizing a novel two-level LSH index based on a new \texttt{adaptivity} metric that is not only asymmetric but also  convertible into Jaccard similarity, and non-similarity-based approaches. 
We conducted extensive comparison experiments on a number of workloads including activity recognition,  image recognition, text classification, entity matching, and natural language processing. The experimental evaluation showed that our proposed asymmetric similarity-based measurement, adaptivity, outperformed symmetric similarity-based measurements and non-similarity-based measurements in most of the workloads. In the future, we will compare these approaches to other approaches such as AutoML and Clipper, which involve labeled target data.

\section{Acknowledgments}
We would like to thank Devshree Chetan Patel, Ratnam Sanjivkumar Parikh, and Dunchuan Wu, for their help to this work.

\bibliographystyle{ACM-Reference-Format}
\bibliography{refs}

\end{document}

%% file: intro.tex
\section{Introduction}
\label{sec:intro}

Reusing a pre-trained model rather than finetuning or retraining can avoid the substantial efforts required for collecting and annotating training data, and will greatly simplify the model deployment costs~\cite{yang2017deep}. However, when there are multiple candidate models, the prediction of the serving accuracy of a model in a target query dataset is inherently a complex problem. It is related to a number of factors, such as the model architecture, initial weights, hyper-parameters, the similarity and adaptivity between the source training dataset and target query datasets~\cite{gonzalez2015optimal}, etc. In this work, we focus on a broad class of model versioning scenarios, where the users need to choose a few models from a large repository of models that have similar architectures~\cite{olston2017tensorflow}.  

\noindent
\textbf{Motivating Scenario 1. Marketplace-level Forecasting at Uber.} They forecast supply, demand, and other quantities in real time for hundreds of cities across the globe. Uber is operating in many cities across the world, and different cities may pose different geospatial characteristics~\cite{sun2020gallery}. Besides, the Uber business might be at different growth stages for different cities. Therefore, they divide the problem spatially by city, training a model instance for each city-quantity combination using the same model architecture. In addition, they frequently update the models through retraining, when they detect model performance
degradation due to the changing market conditions, and they need
to independently trigger the retraining of the models for a city. All of these lead to thousands of or even more model versions~\cite{sun2020gallery}.

\noindent
\textbf{Motivating Scenario 2. Downtime Prediction at Manufacturers.} A steel plate manufacturer~\cite{weber2020model} owns several plants in different countries. The data of machines from different locations are collected centrally. This includes time-series data, such as temperature, humidity, and vibration data, as well as high-resolution image data from cameras. Bob wants to create a machine learning model that predicts the downtimes for a specific machine type. However, the failure situation is changing from time to time, therefore Bob retrains and redeploys the model from time to time to keep it up-to-date, thus leading to many different versions.

\vspace{5pt}
Model versioning has motivated a number of model management platforms, such as Gallery~\cite{sun2020gallery}, ModelDB~\cite{vartak2016modeldb}, ModelHub~\cite{miao2017modelhub}, Metadata Tracking~\cite{schelter2017automatically}, MLFlow~\cite{chen2020developments}, TensorFlow Hub~\cite{tensorflowhub}, PyTorch Hub~\cite{pytorchhub}, and DLHub~\cite{dlhub}. However, the discovery of the proper model versions for serving is mostly through the intuition of domain experts and trial-error processes. 
Most of these platforms provide a model search tool based on the model metadata as shown in Tab. ~\ref{tab:model-metadata}.

\begin{table}[htb]
        \centering
        \scriptsize
        \caption{\label{tab:model-metadata} Model search based on model metadata}
        \begin{tabular}{|c|c|} 
             \hline
            Search Category        & Filter Examples                                                  \\ \hline \hline
            Publication Schemas    & author, date, description                                        \\ \hline
            Model Information      & algorithm, software version, network architecture, dependency    \\ \hline
            Development Provenance & versions, contributors                                           \\ \hline
            Training Information   & training datasets, parameters                                    \\ \hline
            Performance            & accuracy, recall, precision, evaluation dataset, evaluation date \\ \hline
        \end{tabular}
\end{table}

However, none of the above information can directly tell \textit{which version(s) of a model should be selected for inference on a new dataset}, which is exactly the problem that we will address in this study. We observed that existing model serving platforms lack an efficient mechanism to search for one or a few model versions that will have the best inference accuracy on the target query dataset. Instead, users need to perform a text-based search using keywords and scrutinize each search result by deploying and testing the model. Such human-centered  model selection based on trial and error is inefficient, which delays the model deployment and incurs significant human costs. To alleviate such overheads and human costs, it is urgent to automate this process. 

Existing neural architecture search or AutoML~\cite{thornton2013auto,feurer2020auto,ledell2020h2o,olson2016tpot} cannot solve the problem as well. That's because these techniques are designed for the training stage rather than the deployment stage. At the deployment stage, we do not have the labels of the testing data, so it is not practical to retrain a model from scratch. Also, existing AutoML techniques do not consider how to reuse pre-trained models, which are only available after the training stage.

In this work, we proposed and compared multiple model search strategies that are designed for the deployment stage, as follows:

\vspace{5pt}
\noindent
\textbf{1. Symmetric data similarity-based approach.} A basic assumption in learning theory is that the training and testing sets are drawn from the same probability distribution~\cite{gonzalez2015optimal}. If the probability distribution of the testing data is different from the training dataset, the pattern induced from the training dataset may not be relevant to the testing data, and thus such difference will result in poor prediction accuracy~\cite{lu2018learning}.

Therefore, if the pre-trained model's source domain (i.e., the domain of the training data) has similar probability distribution with the target domain (i.e., the domain of the testing data), the pre-trained model may achieve good accuracy in the target domain. This motivates a nearest neighbor search approach that selects the models, of which the training datasets are most similar to the target query dataset. 

\begin{figure}[h]
\centering
   \includegraphics[width=3.4in]{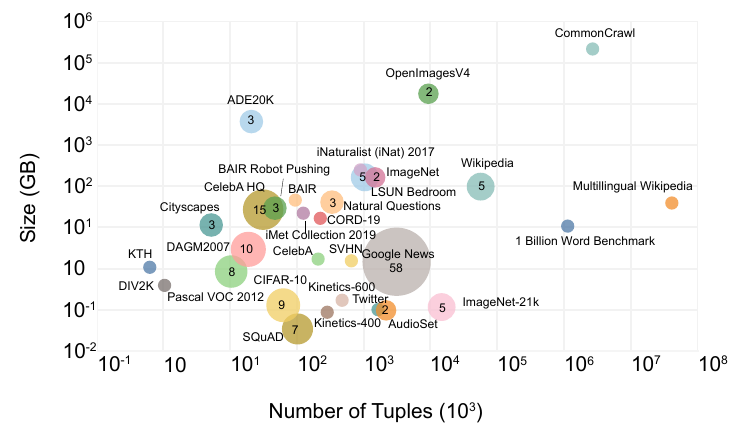}
\caption{\label{fig:tfhub}
TensorFlow Hub Models. The value and size of the circle denote the number of models trained on the dataset.}
\end{figure}

However, given a large number of models and large size of training data (as illustrated in Fig.~\ref{fig:tfhub}),  pair-wise comparison of all models' training datasets to the target query dataset is not practical for high-frequency model search requests~\cite{sun2020gallery}. For example, existing studies~\cite{slow-page, uber-open-summit} show that if search response times increase from $1$ to $4$ seconds, user experience, as well as conversion rates, drop sharply. 

To reduce the pair-wise comparison overhead, we consider leveraging Locality Sensitive Hash (LSH), which is a popular technique to solve nearest neighbor problems. %
Recently, the LSH for measuring the similarity of probability distributions, such as JS-divergence is studied~\cite{chen2019locality, mao2017s2jsd}, which is applied in this work and called JSD-LSH.

\vspace{5pt}
\noindent
\textbf{2. Asymmetric data similarity-based approaches.}
The inference accuracy is asymmetric. For example, a dataset $\mathbf{A}$ may contain only a subset of instances of another dataset $\mathbf{B}$; then supposing we have a model $M_{\mathbf{A}}$ that is trained on $\mathbf{A}$, and a model $M_{\mathbf{B}}$ that is trained on $\mathbf{B}$, the accuracy of $M_{\mathbf{A}}$ on $\mathbf{B}$, and the accuracy of $M_{\mathbf{B}}$ on $\mathbf{A}$ could be very different. Because of such asymmetry, it is problematic to simply apply the symmetric similarity measurement to the features of the training dataset and the features of the target query dataset. 

To address the problem regarding the asymmetric serving accuracy on the swapping of source and target domains, we propose a novel similarity measurement called \texttt{adaptivity}. It first requires partitioning the training dataset and the target query dataset into multiple partitions of the same size respectively. Then, the \texttt{adaptivity} is computed as the set containment of the training dataset to the target query dataset, which is defined as the ratio of the number of similar partitions to the number of the partitions in the target query dataset. 
A great benefit of the \texttt{adaptivity} metric is that it can be converted to Jaccard similarity~\cite{broder1997resemblance}, which motivates us to use the Minwise LSH to accelerate the pair-wise set containment computations. However, the problem is that Minwise LSH is designed for indexing high-dimensional vectors, but in our case, each dataset is a vector of partitions, and it is hard to directly derive the Minwise LSH for such a complicated vector, lacking an efficient comparator for two partitions. To address the problem, we further propose a novel two-level LSH index. The top level is the aforementioned Minwise LSH index, and the bottom level is the aforementioned JSD-LSH.
Utilizing the JSD-LSH, each partition is converted to a JSD-LSH signature, and two partitions have the same signature if they have similar probability distributions (i.e., small JS-divergence). By seeing each signature band as a value, a dataset is abstracted as a vector of values, from which Minwise LSH can be easily computed for estimating the \texttt{adaptivity}.

\vspace{5pt}
\noindent
\textbf{3. Non-similarity-based Approaches.} When the schemas of the training datasets of candidate models are inconsistent with target datasets, we need to compute the overlapping of the source and target datasets at runtime, because such overlap cannot be predicted and precomputed before an ad-hoc model search task is issued. In such a situation, we need to store the training dataset of each candidate model to facilitate such runtime overlap computing, which brings significant storage overheads and complexity in managing such datasets. Therefore, in this study, we also considered two non-similarity-based approaches to avoid such issues. The first approach leverages the predictions of all candidate models to vote for the label of a targeting sample and select the model whose predictions are mostly voted as the best model. While this approach shares some similarities with ensemble inferences, its purpose is to select the best model for deployment, rather than making inferences. The second approach simply selects the model that has the best accuracy in its source domain from which it is trained, regardless of the target domain.

\vspace{5pt}
\noindent
\textbf{Our Key Contributions} are summarized as follows:

\vspace{3pt}
\noindent
(1) In this work, we identified the problem of searching models for serving a target dataset from a set of candidate models that are trained using a  similar methodology with different training data. Our work will greatly enhance the capability of existing model repositories and improve the productivity of the model reuse process, which is urgently needed in production. 

\vspace{3pt}
\noindent
(2) We proposed various model search methodologies that do not rely on any labeled datasets, including the approaches based on various symmetric measurements about the similarity between the source domain and target domain, the approaches based on asymmetric similarity measurements, and the non-similarity-based approaches such as the voting approach, and the approach solely based on source accuracy.

\vspace{3pt}
\noindent
(3) We implemented the proposed approaches and conducted extensive experiments to compare their effectiveness and efficiency using five different applications: activity recognition, image recognition, text classification, entity matching, and natural language processing (NLP). Our study has shown that our proposed asymmetric similarity-based measurement, adaptivity, outperformed symmetric similarity-based measurements and non-similarity-based measurements in most of the workloads.

%% file: insights.tex
\subsection{Key Observations}
We summarized our key observations of various model selection strategies at deployment time, not using any labeled data, as follows:

\vspace{3pt}
\noindent
(1) Asymmetric similarity-based measurements such as adaptivity outperformed symmetric similarity-based measurements such as JS-divergence and L2 distance for most of the workloads except the text classification workload.

\vspace{3pt}
\noindent
(2) Similarity-based measurements, such as adaptivity, JS-divergence, and L2 distance, achieved a significantly lower error rate than non-similarity-based measurements in most of the workloads except the text classification workload. 

\vspace{3pt}
\noindent
(3)  The similarity-based measurements require the precomputation of LSH signatures. The source-accuracy-based measurement requires the collection of the training accuracy of candidate models. The voting approach does not require any preprocessing, but it incurs the most runtime computational overheads for running each candidate model on the target dataset. 

\vspace{3pt}
\noindent
(4)  When the schemas of candidate models are inconsistent with the schema of the target dataset (e.g., using a different dictionary), we need to store the training datasets of candidate models for similarity-based measurements to compute the overlapping features between source training datasets and the target inference datasets at runtime, which incurs significant storage overhead and management complexity. The non-similarity-based measurements will not have such a problem. 

%% file: background.tex
\section{Background}
\label{sec:background}
\subsection{Jensen-Shannon (JS) Divergence}

JS-divergence~\cite{lin1991divergence} is a measurement of the similarity between two probability distributions, which is a symmetric metric derived from the asymmetric Kullback-Leibler divergence~\cite{kullback1951information}.

Let $P$ and $Q$ be two probability distributions associated with a common sample space $\Omega$, and let $M = (P+Q)/2$. The JS-divergence is defined by:

\begin{equation}
    D_{JS}(P \parallel Q) = \frac{1}{2}D_{KL}(P \parallel M) + \frac{1}{2}D_{KL}(Q \parallel M)
\end{equation}

Here, $D_{KL}$ denotes the Kullback-Leibler divergence, which is defined as the following for discrete distributions:

\begin{equation}
    D_{KL}(P \parallel Q) = \sum_{\textbf{x} \in \Omega}{P(\textbf{x} ) \log (\frac{P(\textbf{x} )}{Q(\textbf{x})})},
\end{equation}

\noindent
and as following for continuous distributions:

\begin{equation}
    D_{KL}(P \parallel Q) = \int_{\Omega}{ P(\textbf{x} )\log (\frac{P(\textbf{x} )}{Q(\textbf{x} )})\dif{\textbf{x} }}
\end{equation}

We focus on using JS-divergence to calculate the similarity of two datasets in this paper. The reasons are: First, JS-divergence is a widely used similarity metric for probability distributions~\cite{lee2000measures}. Second, it is easier to find LSH schemes for JS-divergence, compared to other similarity measurements of probability distributions~\cite{mao2017s2jsd}.

While other metrics such as Maximum Mean Discrepancy (MMD) could measure the domain adaptivity more accurately, it requires an optimization process and significantly higher computational costs, as illustrated in Tab.~\ref{tab:latency}.

\begin{table}[h]
\centering
\scriptsize
\caption{\label{tab:latency} Average latency comparison on Activity Recognition (Sec.~\ref{workloads}) datasets (Unit: seconds).}
\begin{tabular}{|r|r|r|r|} \hline
Jaccard similarity&KL-divergence&JS-divergence (w/o LSH)&MMD\\\hline \hline
$379$&$29$&$87$&$6451$\\ \hline
\end{tabular}
\end{table}

\subsection{Locality Sensitive Hashing (LSH)}
LSH was developed for the general approximate nearest neighbor search problem~\cite{indyk1998approximate}. It requires a family of LSH functions, each of which is a hash function whose collision probability increases with the similarity of the inputs. The $(r_1, r_2, p_1, p_2)$-sensitive LSH family is formally defined in Definition.~\ref{def1}.

\begin{definition}
\label{def1}
Let $\mathcal{F}=\{h: M \rightarrow U\}$ be a family of hash functions for distance measurement $D$. $\mathcal{F}$ is $(r_1, r_2, p_1, p_2)$-sensitive  $(r_1 < r_2$ and $p_1 > p_2)$, if $\forall p, q \in M$, it satisfies that: (1) if $D(p, q) \leq r_1$, we have $Pr[h(p) = h(q)] \geq p_1$; (2) if  $D(p, q) \geq r_2$, we have $Pr[h(p) = h(q)] \leq p_2$.
\end{definition}

LSH is first proposed by Indyk et al.~\cite{indyk1998approximate} for measuring the Hamming distance in a $d$-dimensional Euclidean space, which requires the data to be vectors with fixed dimensions. MinHash~\cite{broder1997resemblance} is a widely used family of hash functions for Jaccard similarity. SimHash is an LSH scheme for Cosine distance~\cite{charikar2002similarity}. Both Minwise LSH and SimHash are only applicable to high-dimensional binary vectors or sets of values (without fixed dimensions), and MinHash is usually considered to be more computationally efficient than SimHash~\cite{zhu2016lsh}. 

\subsubsection{Minwise LSH}
\label{sec:minhash}
Minwise LSH~\cite{broder1997resemblance} was proposed as an index for estimating the Jaccard similarity between two sets of values. It converts each set into a MinHash signature using a family of Minwise hash functions. Each Minwise function hashes each value in the set into an integer and returns the minimum hash value. 

Given a Minwise hash function $h_{min}$, it has been proved that the probability of the two minimum hash values, which are observed over two sets of values $X$ and $Y$, equal to each other, is the Jaccard similarity of $X$ and $Y$:
$P[h_{min}(X) = h_{min}(Y)] = jaccard(X, Y) = \frac{X \cap Y}{X \cup Y}$~\cite{broder1997resemblance}.
Furthermore, usually, a family of $K$ Minwise hash functions are applied to a set of values, generating $K$ minimum hash values as a signature. Given the signatures of $X$ and $Y$, Jaccard similarity can be estimated as the ratio of the number of collisions in the corresponding minimum hash values to $K$. In practice, for convenience, the $K$ values are further partitioned into $L$ bands, so that the Jaccard similarity can be estimated as the ratio of the number of bands from the two sets that are equivalent to the total number of bands. The lookup of matching bands can be accelerated by storing the Minwise signatures of all candidates corresponding to each band into a hash table.

\subsubsection{LSH for JS-divergence}
\label{sec:jsd-lsh}
As mentioned, the LSH schemes for probability distributions are recently studied~\cite{chen2019locality, mao2017s2jsd}. S2JSD-LSH~\cite{mao2017s2jsd}  provides LSH functions for a measurement that approximates the square root of two times the JS-divergence. Unfortunately, their LSH design doesn't provide any bound on the actual JS-divergence. Then Chen et al.~\cite{chen2019locality} propose a new LSH scheme for the generalized JS-divergence through the approximation of the squared Hellinger distance, which is proved to be bounded with the actual JS-divergence and thus used in this work, as defined in Eq.~\ref{eq:eq1}:

\begin{equation}
h_{\textbf{a},b}=\lceil{\frac{\textbf{a} \cdot \sqrt{P}+b}{r}}\rceil
\label{eq:eq1}
\end{equation}

\noindent
where $P$ is a probability distribution in the sample space $\Omega$, $\textbf{a} \sim \mathcal{N}(0, I)$ is a $|\Omega|$-dimensional standard normal random vector, $\cdot$ represents the inner product operation, $b$ is a random variable uniformly distributed on $[0, r]$, and $r$ is a positive real number. This approximation is proved to be lower bounded by a factor $0.69$ for the JS-divergence~\cite{chen2019locality}. 

 \eat{
\subsection{Transfer Learning}

}

%% file: solution.tex
\section{Problem Definition}
We are given a list of models of similar architecture, each of which is trained using a different training dataset that shares the same schema of features and labels (e.g., generated from similar data pipelines in the same organization). The set of training datasets for these models is denoted as $T=\{t_1, ..., t_k\}$. A training dataset for the $i$-th model is denoted as $t_i=\{<\vec{x}, y>\}$, where each training sample consists of a feature vector $\vec{x}$, and a label $y$. 
Similarly, a target dataset is denoted as $\mathbf{I}=\{<\vec{x'}, y'>\}$, where each testing sample consists of a feature vector $\vec{x'}$, which is known, and a label $y'$, which is unknown and needs to be inferred. The schema of the target dataset is consistent with the training datasets. The question is which model should be selected so that it will achieve the best inference accuracy on $\mathbf{I}$.

While in this work, we focus on the above-formalized problem, our proposed approaches can be extended to handle source and the target datasets that have similar but inconsistent schemas,  by computing their overlapping feature space and applying the proposed approaches to the overlapping features.

\section{Symmetric Similarity Measurements}
In this work, we mainly consider two representative and widely used symmetric similarity measurements: (1) JS-divergence, which models the source and target datasets as two probability distributions and measures their JS-divergence. (2) Euclidean distance (L2 distance), which computes the distance between the centers of the source and target datasets.

In a model search scenario based on JS-divergence, we need first obtain the target dataset, and the (source) training dataset used by each model. Then each source training dataset will be compared to the target dataset. This process not only incurs significant computational overheads but also brings additional complexity, such as storage overheads and privacy issues, for managing the training dataset of each model. 

To address these issues, we can leverage the LSH for JS-divergence, which is recently investigated by Chen et al.~\cite{chen2019locality}. We precompute the JS divergence-LSH (JSD-LSH) signature, as described in Sec.~\ref{sec:jsd-lsh}, for each dataset. Only the JSD-LSH signatures, rather than the training datasets, will be stored in our proposed system, which will not only alleviate the computational overheads but also reduce the storage overheads and avoid privacy issues.

Similarly, when leveraging the Euclidean distance, the center of each training dataset needs to be precomputed and stored. If there exists a large number of models trained on high-dimensional datasets, LSH for Euclidean distance can also be leveraged to accelerate the searching process.

As discussed in Sec.~\ref{sec:background}, we do not consider Jaccard similarity or Minwise LSH, mainly because they are designed for binary vectors or categorical datasets. However, feature vectors used by deep learning are mostly numerical. We do not consider Maximum Mean Discrepancy (MMD), which is a well-known symmetric similarity measurement for two probabilistic distributions, mainly because it incurs significantly higher computational overhead as illustrated in Tab.~\ref{tab:latency}.

\section{Asymmetric Similarity Measurements}
In this section, we first explain why symmetric metrics that measure the similarity of two sets of features cannot be used to estimate whether a model trained in one set will effectively infer the other dataset. Then, to address the issue, we propose a new asymmetric \texttt{adaptivity}. We further propose a novel two-level LSH index framework to accelerate the system-wide computation of \texttt{adaptivity} with theoretical proofs. %

\vspace{5pt}
\noindent
\textbf{Why symmetric similarity metrics will not work for this problem?} A straightforward idea is to simply measure the similarity between the set of features to be inferred, denoted as $F_I=\{(\vec{x'})\}$, and the set of features of each training dataset, denoted as $F_{i}=\{(\vec{x})\}$, and select the model that has the most similar training dataset. However, if the similarity measurement is symmetric, there exists a significant problem that the results will be skewed by the discrepant sizes of the training dataset and target dataset. For example, if $F_I \subset F_i$, the model trained on $t_i$ may have good accuracy on $\mathbf{I}$, because the model has seen all of the samples in $\mathbf{I}$ during the training process. On the contrary, the model trained on $\mathbf{I}$ may have relatively worse accuracy on $F_{i}$, because the model has only seen a portion of samples in $F_{i}$ in training. Therefore, it's more important to measure the ratio of the similar samples to the samples in the target dataset, which is asymmetric.

\subsection{Adaptivity Measurement}
While an asymmetric similarity measurement such as KL-divergence may also address the issue, there is no existing work that has discussed the LSH for measuring KL-divergence. Therefore, the overhead incurred by the pairwise computation of KL-divergence is prohibitive. In this work, we propose a new measurement, called \texttt{adaptivity}, which can be converted to Jaccard similarity, a symmetric similarity measurement, and thus its computation can be accelerated using a novel two-level LSH (See Sec.~\ref{sec:lsh-index}). 

The idea is to first randomly partition $F_i$ and $\mathbf{I}$ into partitions of the same size and then compute a set containment measurement that is defined as the ratio of the number of similar partitions shared by a training dataset and the target inference dataset to the total number of partitions in the target dataset. The formal definition of the metric is as follows:

\begin{definition}
\label{new-metric}
Supposing we have two datasets  $\mathcal{S}_s = \{p_0, ..., p_{m}\}$ and $\mathcal{S}_t = \{q_0, ..., q_{n}\}$, where $p_i (0 \leq i < m)$ and $q_j (0 \leq j < n)$ are partitions that have equivalent sizes (except for the last residue partition), given a threshold $t$, we can join $\mathcal{S}_s$ and $\mathcal{S}_t$ to identify a subset of $\mathcal{S}_t$, denoted as $\mathcal{S}_t^*$, where $\forall q_j \in \mathcal{S}_t^*$, $\exists p_i \in \mathcal{S}_s$, satisfying that $D_{JS}(p_i, q_j) \leq t$.  We denote the total number of partitions in $\mathcal{S}_t^*$ as $|\mathcal{S}_t^*|=l$. Then we can derive a new metric to measure the adaptivity as the set containment of the target dataset ($\mathcal{S}_t$) in the source dataset ($\mathcal{S}_s$), denoted as: $adaptivity(\mathcal{S}_s, \mathcal{S}_t) = \frac{\mathcal{S}_s \cap \mathcal{S}_t}{\mathcal{S}_t} \sim \frac{\mathcal{S}_t^*}{\mathcal{S}_t} = \frac{l}{m}$.
\end{definition}

Based on the definition of the \texttt{adaptivity} metric, we further formulate the problem as illustrated in Def.~\ref{problem1}, which can be easily extended to a nearest neighbor search problem of finding the top-$k$ models that have the highest adaptivity to the target domain.

\begin{definition}
\label{problem1}
We have a database of $n$ pre-trained models (i.e., source tasks in a source domain), represented as $\mathcal{M}=\{M_1, ..., M_n\}$. Each model $M_i \in \mathcal{M}$ has a source domain $\mathcal{S}_i$, which is a set of training instances that have the same types of features. We have a target query dataset, $q$, that is associated with a target domain $\mathcal{T}_q$, which has the same types of features as source domains. Given an \texttt{adaptivity} threshold $t'$, the database should return a set of relevant models, denoted as $\mathcal{M}_q \subset \mathcal{M}$, so that $\forall M_i \in \mathcal{M}_q, adaptivity(\mathcal{S}_i, \mathcal{T}_q) \geq t'$. 
\end{definition}

\noindent
\textbf{Justification of the Adaptivity Metric.} It is a common practice of leveraging partitioning to improve the accuracy of LSH in an asymmetric scenario~\cite{zhu2016lsh}. Based on partitioning, the set containment measurement is asymmetric, indicating the ratio of the number of matched partitions to the total number of partitions in the \textit{target} dataset. Thus, it can effectively represent the asymmetric relationship between the source domain and the target domain. 
In addition, this set containment measurement and Jaccard similarity can be transformed to each other~\cite{zhu2016lsh}, which facilitates the determination of the threshold $t'$. In addition, because Minwise hash is based on Jaccard similarity (Sec.~\ref{sec:minhash}), we can combine the Minwise hash and the JSD-LSH (for measuring ) to facilitate the computation of the adaptivity metric, as described in detail in the next section.

\subsection{Novel Two-Level LSH Index}
\label{sec:lsh-index}
In this section, in order to accelerate the computation of the adaptivity, we propose a novel two-level LSH index that uses JSD-LSH~\cite{chen2019locality} for measuring the similarity between partitions and uses Minwise LSH~\cite{broder1997resemblance} for measuring the adaptivity from the source dataset to the target dataset.%
\eat{
\begin{figure}[h]
\centering
   \includegraphics[width=3in]{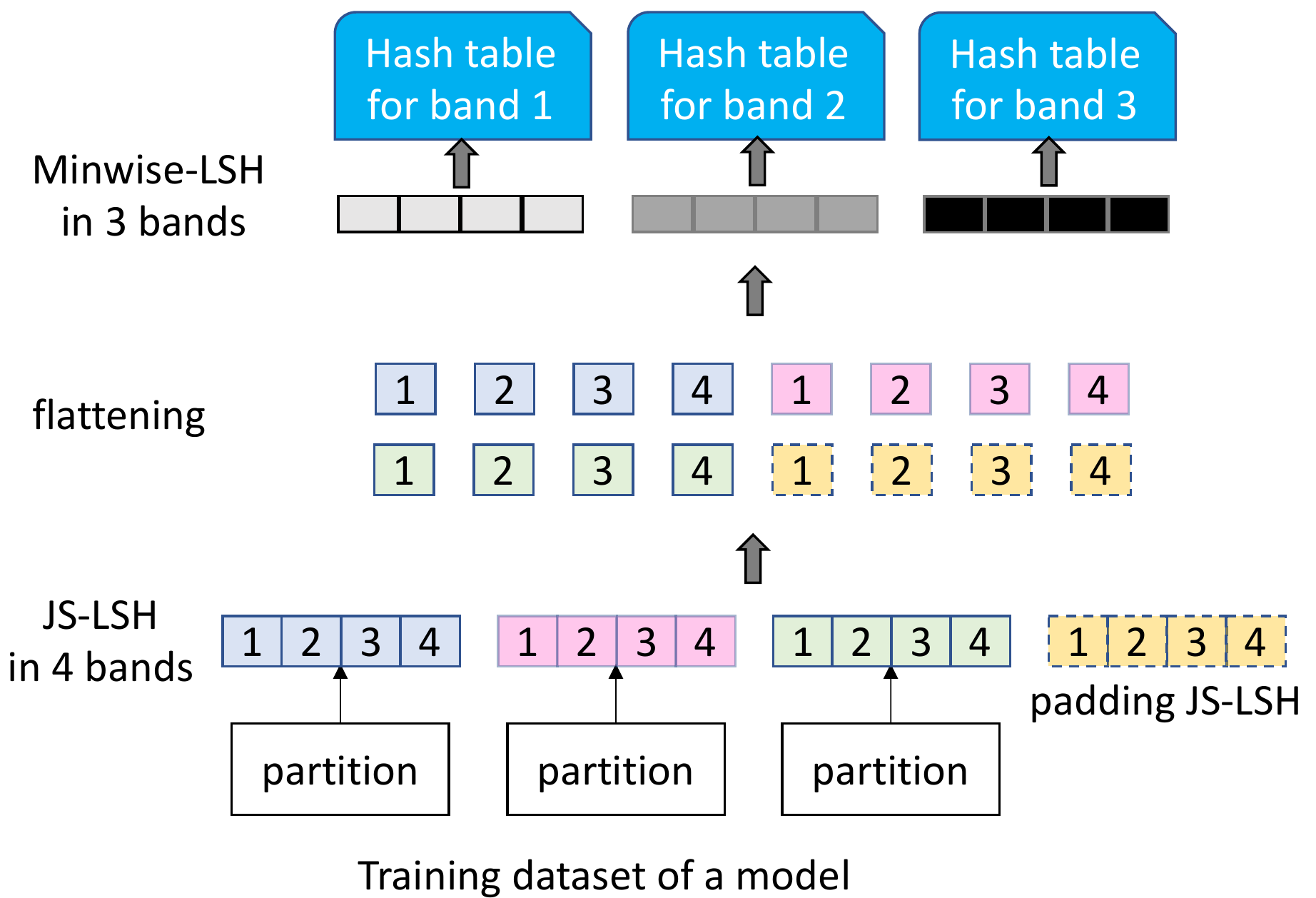}
\caption{\label{fig:2-level-lsh}
Illustration of the two-level LSH index}
\end{figure}
}
\subsubsection{JSD-LSH}
\label{sec:jsd-lsh1}
We first consider the sensitiveness of JSD-LSH applied to each partition. The hash functions $\{h_{\textbf{a},b}\}$, defined in Eq.~\ref{eq:eq1}, form a $(t, c^2\frac{U(\lambda)}{L(\lambda)}t,$ $e_1, e_2)$-sensitive family for the generalized JS-divergence with parameter $\lambda$ (it reduces to usual JS-divergence if $\lambda=0.5$)~\cite{chen2019locality}. \eat{Here, $t>0$,  $c= \norm{\sqrt{P}-\sqrt{Q}}_2$, $U(\lambda)$ is the upper bound function, $L(\lambda)$ is the lower bound function, $e_1=p(1)$, and $e_2=p(c)$, where $p(u)=\int_{0}^{r}{\frac{1}{u}f_2(x/u)(1-x/r)\dif{x}}$ and $f_2$ is the probability density function of the absolute value of the standard norm distribution.}
This means,  we can leverage $Pr[h(p_i)=h(q_j)]\geq e_1$, to determine whether $D_{JS}(p_i, q_j) \leq t$. We simply denote this mapping relationship as $e_1 = g(t)$. Please see ~\cite{chen2019locality} for the proof and more details. In practice, we use a family of $K$ JSD-LSH functions by randomly choosing the hyperparameters $\textbf{a}$ and $b$. Then we will obtain $K$ values for each partition by applying the $K$ functions. These $K$ values are further partitioned into $L$ bands. We estimate the probability that $p_i$ and $q_j$ are similar (i.e., $Pr[h(p_i)=h(q_j)]$) as the ratio of the matched JSD-LSH bands to $L$.  If we see each JSD-LSH signature as a set of JSD-LSH bands, the $Pr[h(p_i)=h(q_j)]$ is actually the Jaccard similarity of the two sets. This means $jaccard(h(p_i), h(q_j)) \geq g(t)$, which connects the value of the JS-divergence threshold $t$ as defined in Def.~\ref{new-metric} and the Jaccard similarity threshold $t'$ defined in Def.~\ref{problem1}, as we explained in the next section.

\subsubsection{Minwise LSH}
\label{sec:translation}
Now we consider the adaptivity threshold $t'$, given source dataset $\mathcal{S}_s = \{p_0, ..., p_{m}\}$, and target dataset $\mathcal{S}_t = \{q_0, ..., q_{n}\}$, Jaccard similarity is defined as $jaccard(\mathcal{S}_s, \mathcal{S}_t) = \frac{{S}_s \cap {S}_t}{{S}_s \cup {S}_t}= \frac{l}{m+n-l}$, which means
$jaccard(\mathcal{S}_s, \mathcal{S}_t) = \frac{adaptivity(\mathcal{S}_s, \mathcal{S}_t)}{\frac{n}{m}+1-adaptivity(\mathcal{S}_s, \mathcal{S}_t)}$. 

Therefore, if $adaptivity(\mathcal{S}_s, \mathcal{S}_t) \geq t'$, it means the Jaccard similarity between source domain partitions and target domain partitions satisfies
$jaccard(\mathcal{S}_s, \mathcal{S}_t) \geq \frac{t'}{\frac{n}{m}+1-t'}$. We leverage a family of Minwise hash functions as described in Sec.~\ref{sec:minhash} to generate a Minwise hash signature for each dataset (i.e., regarded as a set of partitions), with each partition represented as a JSD-LSH signature. In addition, as mentioned at the end of Sec.~\ref{sec:jsd-lsh}, the equivalence of $p_i$ and $q_j$ is determined by $jaccard(h(p_i), h(q_j)) > g(t)$, where $h(x)$ represents the JSD-LSH function that maps $x$ to a set of $L$ bands, which is denoted as $h(x)=\{b_1, ..., b_L\}$. Therefore, if $g(t)=1/L$, we can simplify the computation of such recursive Jaccard similarity measurement based on below equation that helps us to derive an upper-bound of the adaptivity metric:

\begin{align*}
&jaccard(\mathcal{S}_s, \mathcal{S}_t) \leq L \times jaccard(\cup_{p_i}{\{b_i|b_i\in p_i\}}, \cup_{q_j}{\{b_j|b_j\in q_j\}})
\end{align*}

\noindent
\textbf{Proof.} The intuition is that if two sets of items (i.e., each item is a JSD-LSH signature computed on a partition) are similar, then if we split each item into sub-items (e.g., each sub-item is a band in a JSD-LSH signature), the two sets of sub-items are still similar to each other, and vice versa. 
More formally, supposing $\exists A \subset \mathcal{S}_s$ satisfying $\forall p_i \in A, \exists q_j \in \mathcal{S}_t$, and $jaccard(h(p_i), h(q_j))\geq 1/L$, we have $jaccard(\mathcal{S}_s, \mathcal{S}_t)=\frac{|A|}{|\mathcal{S}_s\cup\mathcal{S}_t|}$.
Additionally, $jaccard$ $(h(p_i), h(q_j))\geq 1/L$ indicates that $\exists b_i \in p_i$ and $b_j \in q_j$, satisfying $b_i=b_j$. 
Therefore, $jaccard(\cup_{p_i}{\{b_i|b_i\in p_i\}}, \cup_{q_j}{\{b_j|b_j\in q_j\}})\geq \frac{|A|}{|\mathcal{S}_s\cup\mathcal{S}_t|\times L}$ Therefore, $jaccard(\mathcal{S}_s, \mathcal{S}_t) \leq L \times jaccard(\cup_{p_i}{\{b_i|b_i\in p_i\}}$, $ \cup_{q_j}{\{b_j|b_j\in q_j\}})$, and we can use the latter as an upper-bound estimation of the former metric.

Based on the approximation, a JSD-LSH signature is first generated for each partition and each signature is further split into $L$ bands. Then the set of partitions in a dataset will be flattened into a set of JSD-LSH bands.
Given user-specified adaptivity threshold $t'$, we can decide whether two sets have their Jaccard similarity higher than $\frac{t'}{\frac{n}{m}+1-t'}$ times a factor of $L=\frac{1}{g(t)}$ by comparing their Minwise hash signatures as described in Sec.~\ref{sec:minhash}.

One potential problem is that the value of $n$ is probably not a constant shared by all training datasets. To address the problem, we leverage a padding strategy as in asymmetric Minwise LSH~\cite{shrivastava2015asymmetric}. We select an upper bound of $n$, denoted as $n_u$, then we pad every training dataset to have $n_u$ partitions by appending ($n_u-n$) JSD-LSH signatures for non-existing partitions. The values in these padding signatures will not exist in real signatures, so the padding will not affect the precision, while it may lead to false negatives. \eat{But we argue that, in a model discovery scenario, the users prefer to be returned with a small set of relevant models, instead of a large set of models that contain false positives. On the contrary, if the recall is more important than precision, LSH-ensemble~\cite{zhu2016lsh} should be applied. In addition, in our targeting scenario, LSH-ensemble requires further partitioning the set of JSD-LSH signatures, and finetune the sizes of partitions to achieve a reasonable recall, which may introduce significantly more complex compared to the padding strategy.}

The algorithm based on the two-level LSH index to address the problem formulated in Def.~\ref{problem1} is illustrated in Alg.~\ref{alg:search1}. %

\begin{algorithm}[!ht]\small
\caption{\bf Model discovery algorithm}
\label{alg:search1}
\begin{algorithmic}[1]
\STATE INPUT1: $\mathcal{T}_q$  (query dataset, i.e., the target dataset)
\STATE INPUT2: $t'$ (the adaptivity threshold)
\STATE INPUT3: $L$ (the number of bands in each JSD-LSH hash)
\STATE INPUT4:$L_m$ (the number of bands in each Minwise hash)
\STATE INPUT5: $\{hashmap_i\} (1\leq i\leq L_m)$ (one hashmap for each band of the Minwise-LSH.) 
\STATE OUTPUT: $\mathcal{M}_q$ %
\STATE $m \leftarrow |\mathcal{T}_q|$ %
\STATE $t^* \leftarrow L\times \frac{t'}{\frac{u}{m}+1-t'}$
\STATE $Q \leftarrow partition(\mathcal{T}_q)$ %
\STATE $S \leftarrow \phi$
\FOR{$q \in Q$}
    \STATE $S \leftarrow S \cup JSDLSH(q)$
\ENDFOR
\STATE $S' \leftarrow MinwiseLSH(S)$ \COMMENT{$S'$ consists of $L_m$ bands}
\STATE $\mathcal{M}, \mathcal{M}_q \leftarrow \phi$
\FOR{$s_i \in S'$}
   \STATE $\mathcal{M}_i \leftarrow hashmap_i.lookup(s_i)$
   \STATE $\mathcal{M} \leftarrow M \cup \mathcal{M}_i$
\ENDFOR
\FOR{$M_i \in \mathcal{M}$}
   \IF{$\frac{\mathcal{M}.count(M_i)}{L_m} > t^*$}
      \STATE $\mathcal{M}_q \leftarrow \{m\} \cup \mathcal{M}_q$
   \ENDIF
\ENDFOR
\RETURN $\mathcal{M}_q$
\end{algorithmic}
\end{algorithm}

\section{Non-Similarity-Based Approaches}

\vspace{5pt}
\noindent
\textbf{Motivation.} When the schemas of the training datasets of candidate models are consistent with the target dataset, the similarity-based approaches can work efficiently, because each dataset can be abstracted as an LSH signature, and only the signature needs to be stored and queried at runtime. However, if we extend the work to a situation where the schemas of the training datasets are not consistent with the target dataset, we need to first identify the overlapping features that are shared by the target dataset and the training dataset of each candidate model. Then we apply the similarity measurements to the overlapped features. Identifying the overlapping features for each candidate model not only requires a lot of computation time but also requires storing all training datasets of candidate models. This may be possible in certain scenarios, e.g., models are natively served from a data management system~\cite{zhou2022serving, zou2020lachesis, yuan2020tensor, jankov2019declarative, zou2019pangea}. However, for general use cases, it is impractical to request users to make their training datasets available for model searching and deployment.

In addition, we found that the similarity measurements only work well when some of the training datasets are correlated with the target datasets. If the training datasets and the target datasets are drawn from two different domains (e.g., the corpus of clinical healthcare records vs. the corpus of movie reviews), similarity-based measurements are not very helpful.

Therefore, to address these limitations of the similarity-based model search approaches, in this study, we also considered two non-similarity-based approaches that do not rely on the training datasets of candidate models, which are described as follows.

\vspace{5pt}
\noindent
\textbf{1. A Voting Approach.} This approach has two phases: \textit{voting} and \textit{selection}. 
At the beginning of the voting phase, every candidate model has a credit counter. Then, during the voting phase, each candidate model will make a prediction for each sample in the target dataset. For extensions to the situation where the target dataset's schema does not match the candidate model's input feature format, a preprocessing step will be required to convert the target feature to each model's required input format via padding, truncating, reordering, etc.  

For each target sample, the prediction that is mostly voted by the candidate models becomes the winner, and all candidate models that make a winning prediction will have their credit counter incremented by one. 

In the selection phase, the model that has the maximal number of credits is selected for serving the targeting dataset. While this approach shares some similarities with ensemble inferences, its purpose is to select the best model for deployment, rather than making inferences directly.

\vspace{5pt}
\noindent
\textbf{2. An Approach based on Source Accuracy.} Though the voting approach does not require the storage of the training datasets of the candidate models, the voting process is computation-intensive and if there exists schema inconsistency, the preprocessing of target datasets for voting is difficult to be automated. These limitations motivate a much simpler approach, which selects the model that has the best source accuracy, which is defined as the accuracy of the model in its source domain from which it is trained (i.e., training accuracy),  regardless of the target domain. This approach requires collecting the source accuracy information for each candidate model. 

\vspace{5pt}
\noindent
A \textbf{qualitative comparison} of all proposed model selection approaches is illustrated in Tab.~\ref{tab:strategies} and Tab.~\ref{tab:strategies-1}.

\begin{table} [h]
\centering
\scriptsize
\caption{\label{tab:strategies} High-Level comparison of model search strategies w/o schema inconsistencies}
\begin{tabular}{|p{2cm}|p{1.5cm}|p{1.5cm}|p{2cm}|} \hline
&Runtime Computational Overheads & Storage  Overheads & Preprocessing Overheads \\\hline \hline
JS-Divergence&JSD-LSH computation&LSH signatures&LSH precomputation\\ \hline
L2 Distance&L2-LSH computation&LSH signatures&LSH precomputation\\ \hline
Adaptivity&Two-level LSH computation&LSH signatures&LSH precomputation\\ \hline
Voting&model prediction &none&none\\ \hline
Source Accuracy&top-k filtering&none&Source accuracy info collection\\ \hline
\end{tabular}
\end{table}

\begin{table} [h]
\centering
\scriptsize
\caption{\label{tab:strategies-1} High-Level comparison of model search strategies extended to handle schema inconsistencies}
\begin{tabular}{|p{2cm}|p{2cm}|p{1.5cm}|p{1.5cm}|} \hline
&Runtime Computational Overheads & Storage  Overheads & Preprocessing Overheads \\\hline \hline
Similarity-based approaches&Overlapping \& LSH computation&Training datasets&none\\ \hline
Voting&target feature preprocessing \& Feature preprocessing&none&none\\ \hline
Source Accuracy&top-k filtering&none&Source accuracy info collection\\ \hline
\end{tabular}
\end{table}

\eat{
\subsubsection{Compression based on histogram}

\subsubsection{Compression based on clustering}
}
\eat{
\subsection{Problem Formulation}

\begin{definition}
\label{def3}
\end{definition}

\noindent

\noindent

\noindent

\noindent

\subsection{Overlap Search}
\label{sec:overlap}

\subsection{Model Search}
\label{sec:js}

\eat{
}

\eat{ 
}
}

%% file: evaluation.tex
\section{Empirical Evaluation}
\label{sec:evaluation}
\subsection{Workloads and Datasets}
\label{workloads}
We evaluate our proposed methodology using five workloads.

\noindent
\textbf{1. Activity Recognition.} Human activity recognition (HAR), is to predict the activities (e.g., walking, sitting, running, lying) based on data collected from multiple sensors attached to the human body. HAR is a hot research topic in the pervasive computing area and has been widely applied to indoor localization, sleep state detection, smart home sensing, and virtual reality~\cite{avci2010activity}. In this work, we use three public activity recognition datasets, including OPPORTUNITY~\cite{chavarriaga2013opportunity}, PAMAP2~\cite{reiss2012introducing}, and DSADS~\cite{barshan2014recognizing}.  The OPPORTUNITY dataset is collected from four human subjects executing various activities with sensors attached to more than five body parts. The PAMAP2 dataset is collected from nine subjects performing $18$ activities with sensors attached to three body parts. The DSADS dataset is collected from eight subjects wearing sensors on five body parts.  A dataset is collected for each body part, therefore there are $13$ datasets in total. Each dataset has $81$ features~\cite{wang2018stratified} and $1920$ to $5022$ samples. Then we apply the aforementioned five model search approaches to a series of scenarios where models are trained on each of these tables and given a target dataset collected from a subject on a specific body part, we want to select a model to achieve the best accuracy on the target dataset. 

\noindent
\textbf{2. Image Recognition.}  We randomly sample images from the CIFAR-10 dataset without replacement to create five image datasets with distinct probability distributions. CIFAR-10 is a collection of images with labels of ten classes, with each containing $5,000$ images. Then we make four of the subsets skewed by adding $5,000$ images of the fourth class to the second and the fourth subset; $5,000$ images of the second class to the third subset; $5,000$ images of the eighth class to the third subset.  These five datasets are called as \texttt{Balanced}, \texttt{Skewed-1}, \texttt{Skewed-2}, \texttt{Skewed-3}, \texttt{Skewed-4} correspondingly. To evaluate the effectiveness of our proposed model search strategies for the image recognition scenario, we train ResNet56v1 model on each of the subsets respectively and then search for the best model to be reused on each dataset using different strategies.

\noindent
\textbf{3. Text Classification (Sentiment).} The objective of this task is to classify text as positive or negative sentiment. In this task, we train models on seven different datasets: two $25,000$ randomly sampled IMDB Dataset of Movie Reviews; tweets from Twitter which consist of two datasets of $25,00$ samples each; Financial PhraseBank contains the sentiments for financial news headlines which has $4,846$ samples divided into two datasets; and hate speech tweets dataset which has $4,484$ samples. We train models for these tasks using the pre-trained DistilBERT base model (uncased)~\footnote{https://tfhub.dev/jeongukjae/distilbert\_en\_uncased\_L-6\_H-768\_A-12/1}, %
which contains $6$ layers and $40$\% less parameters than bert-base-uncased. Then for each task, we use one of the seven datasets as the target dataset and try to choose the top- $k$ candidate models to serve on the target dataset using different approaches.

\noindent
\textbf{4. Entity Matching (EM).} This task is to decide \textit{whether two tuples are referring to the same entity}~\cite{li2020deep, zhao2019auto, ebraheem2017deeper, cappuzzo2020creating, fernandez2018seeping}. \eat{Many recent EM tools are based on deep learning~\cite{mudgal2018deep, zhao2019auto, ebraheem2017deeper, zhang2020multi}.} We mainly apply DeepMatcher~\cite{mudgal2018deep}, which is an EM tool based on deep learning, to four datasets~\footnote{https://github.com/anhaidgroup/deepmatcher/blob/master/Datasets.md}: Walmart-Amazon, Abt-Buy, DBLP-Scholar, DBLP-ACM; and $11$ smaller datasets from the Magellan Data Repository~\cite{magellandata}~\footnote{https://sites.google.com/site/anhaidgroup/useful-stuff/data}.   Each task contains training and testing samples collected from two different datasets. For example, IMDB-TMD is to match movie tuples collected from IMDB and TMD respectively. and IMDB-RottenTomatoes is to match movie tuples from IMDB and Rotten Tomatoes respectively. We focus on the model selection scenarios where a dataset is used for query, and a set of models are trained for the EM tasks on the rest of the datasets as candidates, and then we try to select top-$k$ candidate models to serve the query dataset.

\noindent
\textbf{5. Natural Language Processing (NLP).} We are mainly interested in two types of NLP tasks. The first task is to identify whether sentence pairs have equivalent semantic meanings. For this task, we train models on three different datasets: Microsoft Research Paraphrase Corpus (MRPC) which includes $3,549$ samples; Quora Question Pairs (QQP) which consists of $363,192$ samples; and Paraphrase Adversaries from Word Scrambling (PAWS) which has $49,401$ samples. 
The second task is about natural language inference (NLI), which is to identify the textual entailment relationship~\cite{maccartney2009natural} between sentence pairs. Similar to the first task, we train models on three different datasets: Recognizing Textual Entailment (RTE) which has $2,490$ samples; Question NLI, which contains $510,711$ samples; and the SCITAIL dataset including $5,302$ samples, which is an entailment dataset created from multiple-choice science exams and web sentences.
We train models for these tasks using the pre-trained BERT base model~\footnote{https://tfhub.dev/google/bert\_uncased\_L-12\_H-768\_A-12/1}, which contains $12$ layers and $110$ millions of parameters. Then for each task, we use one of the three datasets as the target dataset and try to choose the best model to serve on the target dataset.

\subsection{Evaluation Methodology} 
One major evaluation task is to compare various model searching strategies without labeled data at deployment time. These strategies include:

\noindent
(1) JS-divergence (a special case of adaptivity when each dataset has only one partition), which is a symmetric similarity measurement for two probability distributions, as mentioned earlier in the paper.

\noindent
(2) L2-distance, which is a symmetric similarity measurement to get the center of the instances by averaging all feature vectors for the source and target datasets and then computing Euclidean distance between two centers\eat{~\cite{indyk1998approximate}}; 

\noindent
(3) Adaptivity, which is an asymmetric similarity measurement to compute the adaptivity of the source dataset to the target dataset; 

\noindent
(4) Voting, which is the number of majority-voted predictions (i.e., the number of predictions approved by a majority of all candidate models) made by a candidate model.

\noindent
(5) Source accuracy, which is the accuracy of the candidate model on its training dataset. This information is widely available in most types of model databases and in practice, and users usually rely on this information for manual model discovery.

\eat{For the Activity recognition scenario, we also compare to additional metrics, such as the mean squared error, mean absolute error, and loss when training the candidate model on the training dataset.}

\vspace{5pt}
To compare the effectiveness of these approaches quantitatively, we use the following evaluation metrics:

\noindent
(1) \textbf{Pearson correlation coefficient} (PCC)~\cite{lee1988thirteen}, a measure of the linear correlation coefficient between two random variables, to show the correlation of various metrics to the target accuracy (i.e., the prediction accuracy of the candidate model on the target dataset).

\noindent
(2) \textbf{Top-$k$-error}, is defined as the number of wrong predictions that fail to correctly select the optimal model in the top-$k$ results (ranking of the $k$ selected models is not important here) to the total number of predictions. For example, for the top-$3$ search, if the ground truth is \{C\}, the prediction of \{B, C, D\} is a correct prediction. We use the error rate metric for evaluating the effectiveness of different strategies. 

For the \texttt{adaptivity} measurement, we tune hyperparameters including the value of $r$ for JSD-LSH, the number of hash functions and the number of bands for Minwise Hash and JSD-LSH, and the number of partitions for each test case by searching the combination of hyperparameters in a pre-specified range. 

\subsection{Evaluation Results}

\subsubsection{Activity Recognition}

We first evaluate the Activity Recognition workload for the effectiveness of the proposed metrics.

\noindent
\textbf{Summary.} The comparison of Pearson correlation coefficient values of various metrics, including the adaptivity, the JS-divergence, the L2-distance, and the source accuracy, to the target accuracy are illustrated in Tab.~\ref{tab:comparison-to-model-adaptivity}. It shows that our proposed adaptivity metric and voting achieve a better correlation with target accuracy compared to other metrics. Also, among the four metrics, the source accuracy exhibits the least relevance to the target accuracy, which verifies our assumption that the similarity between the source and the target plays an important role in determining the target accuracy (i.e., model adaptivity to the target domain).
We also compute the top-1, top-2, top-3 error rates, as illustrated in Tab.~\ref{tab:comparison-error-rate}. As mentioned, the Top-$k$ error rate is defined to be the ratio of the number of searches that failed to return all $k$ models correctly to the total number of searches. 
The results show that \texttt{adaptivity} works better than other metrics.

\begin{table}[h]
\centering
\scriptsize
\caption{\label{tab:comparison-to-model-adaptivity} Comparison of Pearson correlation coefficient (PCC) for activity recognition (the best Pearson correlation coefficient values are highlighted in bold).}
\begin{tabular}{|l|p{1cm}|p{1cm}|p{1cm}|p{1cm}|p{1cm}|} \hline
target&pcc of adaptivity&pcc of JS-divergence&pcc of l2-distance&pcc of source-accuracy&pcc of voting\\\hline \hline
dsads\_la&$0.61$&$0.17$&$0.54$&$0.44$&$\textbf{0.91}$\\ \hline
dsads\_ll&$\textbf{0.72}$&$0.40$&$0.68$&$0.40$&$0.68$\\ \hline
dsads\_ra&$0.51$&$0.13$&$0.54$&$0.37$&$\textbf{0.78}$\\ \hline
dsads\_rl&$0.58$&$0.49$&$0.75$&$0.58$&$\textbf{0.87}$\\ \hline
dsads\_t&$0.19$&$0.11$&$0.42$&$0.44$&$\textbf{0.80}$\\ \hline
oppo\_b&$0.89$&$0.66$&$0.77$&$0.45$&$\textbf{0.95}$\\ \hline
oppo\_lla&$0.79$&$0.54$&$0.54$&$0.34$&$\textbf{0.88}$\\ \hline
oppo\_lua&$\textbf{0.91}$&$0.72$&$0.76$&$0.48$&$0.87$\\ \hline
oppo\_rla&$0.76$&$0.54$&$0.52$&$0.28$&$\textbf{0.87}$\\ \hline
oppo\_rua&$0.84$&$0.67$&$0.70$&$0.44$&$\textbf{0.86}$\\ \hline
pamap\_a&$\textbf{0.69}$&$0.68$&$0.54$&$0.24$&$0.27$\\ \hline
pamap\_c&$\textbf{0.77}$&$0.69$&$0.70$&$0.06$&$0.76$\\ \hline
pamap\_w&$\textbf{0.83}$&$0.74$&$0.63$&$0.14$&$0.51$\\ \hline
\end{tabular}
\end{table}

\begin{table}
\centering
\scriptsize
\caption{\label{tab:comparison-error-rate} Comparison of different similarity measurements for activity recognition}
\begin{tabular}{|l|r|r|r|r|r|} \hline
          &adaptivity&JS-divergence&l2-distance&source-accuracy&voting\\\hline \hline
top-1 error&$\textbf{15\%}$&$31\%$&$23\%$&$92\%$&$38\%$\\ \hline
top-2 error&$\textbf{0\%}$&$\textbf{0\%}$&$\textbf{0\%}$&$92\%$&$30\%$\\
\hline
top-3 error&$\textbf{0\%}$&$\textbf{0\%}$&$\textbf{0\%}$&$77\%$&$23\%$\\
\hline
\end{tabular}
\end{table}

\noindent
\textbf{Effectiveness of the Two-Level Index.} Now we compare the accuracy and efficiency of using different adaptivity computation strategies including pair-wise comparison, two-level LSH without flattening, and two-level LSH index with flattening. Flattening refers to the step that flattens a set of partitions into a set of JSD-LSH bands, which is the approach that we leveraged for approximating the recursive Jaccard similarity as justified in Sec.~\ref{sec:translation}. The case where flattening is not applied is equivalent to the case where flattening is applied but the number of JSD-LSH bands $L$ is set to $1$. The results are illustrated in Tab.~\ref{tab:ar-strategy}. Here, for JSD-LSH, we use $r=1.4$, number of bands is $L=200$, number of hash functions is $K=800$. For Minwise hash, we use $256$ Minwise hash functions and use $128$ bands. We also allocate $55$ samples in each partition, we can see that without the flattening step, the accuracy will significantly drop. In addition, the two-level index achieves about $65$ times speedup compared to the pair-wise approach for this scenario. \eat{We expect that with the increase in the number of partitions for larger-scale datasets, and the increase in the number of datasets, the proposed approach will achieve even larger speedup.}

\begin{table}
\centering
\scriptsize
\caption{\label{tab:ar-strategy} Comparison of adaptivity computation strategies for activity recognition}
\begin{tabular}{|l|r|r|r|r|} \hline
          &top-1 error&top-2 error&top-3 error &latency (milli-sec)\\\hline \hline
pair-wise &$\textbf{15\%}$&$\textbf{0\%}$&$\textbf{0\%}$&$24,627$\\ \hline
2-level LSH w/o flattening&$54\%$&$62\%$&$54\%$&$\textbf{377}$\\
\hline
2-level LSH w/ flattening&$31\%$&$8\%$&$\textbf{0\%}$&$\textbf{377}$\\
\hline
\end{tabular}
\end{table}

\noindent

\begin{figure}[h]
\centering
   \includegraphics[width=3.3in]{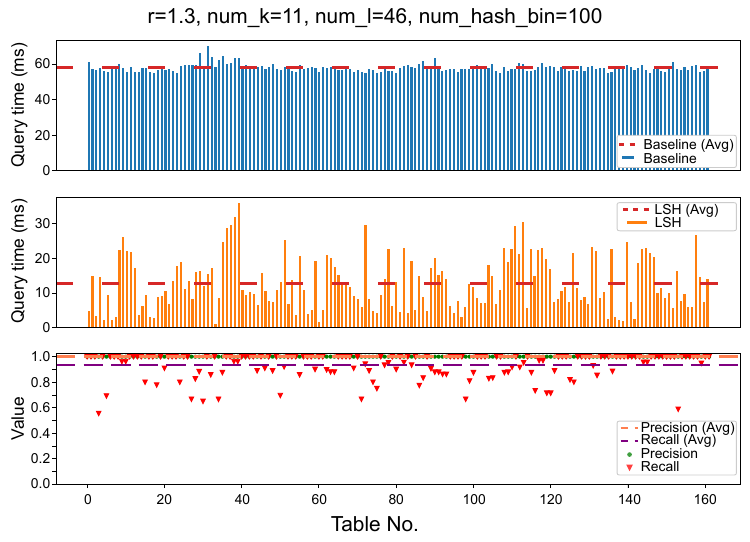}
\caption{\label{fig:ar-comparison}
Comparison of LSH for JS-divergence (denoted as LSH) to a pair-wise computation of JS-divergence (denoted as baseline) for $162$ tables sampled from $13$ activity recognition datasets.}
\end{figure}

\noindent
\textbf{Effectiveness of JSD-LSH.}
To justify the use of JSD-LSH, which is a relatively new LSH function and the first LSH function for indexing JS-divergence with a bound, we compare the accuracy and latency of LSH for JS-divergence to a baseline approach that computes the JS-divergences between the target dataset and each of the source datasets directly without using LSH techniques. For this experiment, we created  $162$ tables by sampling these $13$ activity recognition datasets, so that each table represents samples collected from a body part for a specific subject. The results are illustrated in Fig.~\ref{fig:ar-comparison}, in which each bar represents a test case that uses one table as the query (i.e., target dataset), and the rest of the tables as the sources. In this experiment, for each test, we assume the source datasets that have the JS-divergence with the query dataset smaller than $0.1$ compose the ground truth. We observe that using LSH can significantly accelerate the overall latency required for JS-divergence comparison, and achieve $5$ times speedup on average, while the precision of our proposed approach is $100\%$ and recall is above $93\%$. 

\eat{
\noindent
\textbf{Hyperparameter Tuning for JS-LSH.}
We further evaluate how hyperparameters, such as $r$ (as in Eq.~\ref{eq:eq1}), the number of concatenated hash functions (i.e., $K$) of each band, the number of bands (i.e., $L$), and the number of hash bins for creating probability distribution for comparison, will all affect the accuracy of the JS-divergence computed using LSH compared to the baseline, which are illustrated in Fig.~\ref{fig:ar-tuning}. It shows that it is possible to find a set of parameters that ensure high precision and recall. It also verifies that our approach is optimized for precision and can effectively reduce false positives.

\begin{figure}[H]
\centering
{\subfigure[r]{%
   \label{fig:r}
   \includegraphics[width=3.3in]{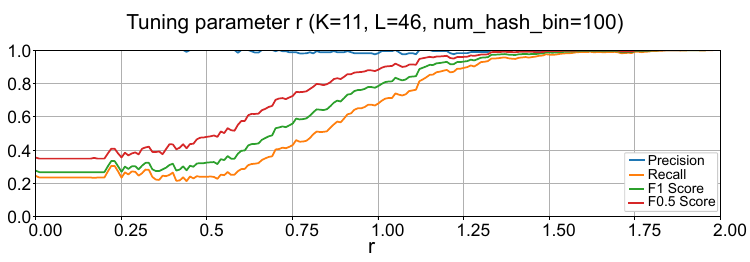}  
}%
\vspace{1pt}
\subfigure[number of concatenated hash functions]{%
  \label{fig:K}
  \includegraphics[width=3.3in]{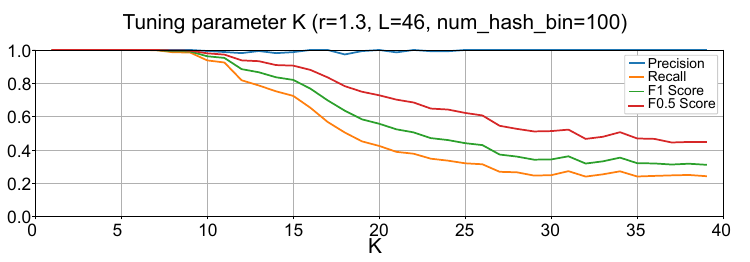}
}%
\vspace{1pt}
\subfigure[number of hash tables]{%
  \label{fig:L}
  \includegraphics[width=3.3in]{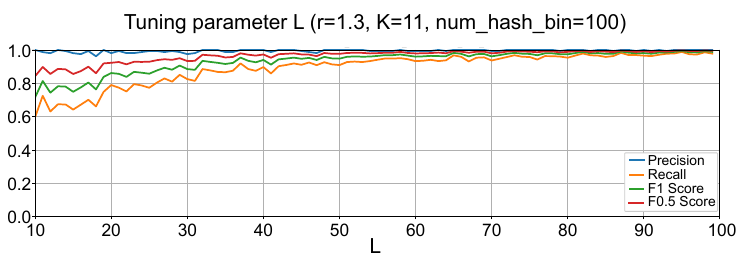}
}%
\vspace{1pt}
\subfigure[number of hash bins]{%
  \label{fig:L}
  \includegraphics[width=3.3in]{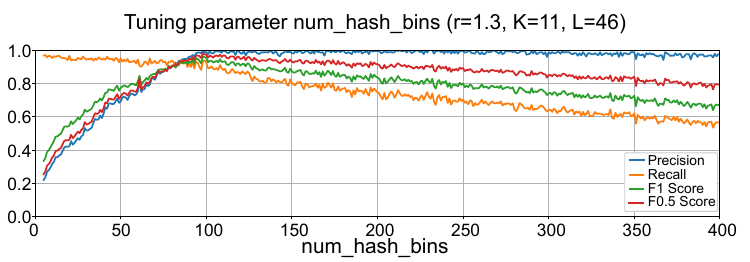}
}
}
\caption{\label{fig:ar-tuning}
Hyperparameter tuning for LSH of JS-divergence.
}

\end{figure}
}

\subsubsection{Image Recognition}
In this scenario, for each CIFAR-10 image, we normalize all pixel values by dividing each value by $255$ and then flatten the image from a tensor of the shape ($32$, $32$, $3$) into a one-dimensional vector of the shape ($1$, $3072$). Then, we compute the \texttt{adaptivity}, JS-divergence, and L2-distance on the flattened dataset. \eat{As mentioned in Sec.~\ref{workloads}, we randomly create five datasets, each having a different probability distribution. Only one dataset, which contains $5000$ images, has a uniform distribution of image classes. Each of the four other datasets consists of $11,000$ images and is skewed in one of the classes. As mentioned in Sec.~\ref{workloads}, we conduct five experiments, and in each experiment, one of the five datasets is considered the target dataset and we try to use different metrics to select a model that is trained on one of the rest of the datasets to serve on the target dataset.}

\noindent
\textbf{Summary.} We evaluate and compare the effectiveness of \texttt{adaptivity}, JS-divergence, L2-distance, and source accuracy for searching for the best model to serve on a target dataset. The Pearson correlation coefficient of each metric to the target accuracy in each experiment is illustrated in Tab.~\ref{tab:comparison-to-model-adaptivity-image}. The top-1 and top-2 error rates for searching for the best model for each experiment are illustrated in Tab.~\ref{tab:comparison-error-rate-image}. We only show top-1 and top-2 error rates, because we have only four candidate models in this case.

\begin{table}
\centering
\scriptsize
\caption{\label{tab:comparison-to-model-adaptivity-image} Comparison of Pearson correlation coefficient for image recognition (the best Pearson correlation coefficient values are highlighted in bold)}
\begin{tabular}{|l|p{1cm}|p{1cm}|p{1cm}|p{1cm}|p{1cm}|p{1cm}|} \hline
target&pcc of adaptivity&pcc of JS-divergence&pcc of l2-distance&pcc of source-accuracy&pcc of voting\\\hline \hline
Balanced&$\textbf{0.93}$&$0.92$&$0.62$&$0.80$& $0.88$\\ \hline
Skewed-1&$\textbf{0.86}$&$0.53$&$0.52$&$0.78$& $0.48$\\ \hline
Skewed-2&$0.39$&$\textbf{0.70}$&$0.66$&$0.07$& $0.22$\\ \hline
Skewed-3&$0.25$&$0.06$&$0.18$&$\textbf{0.73}$& $0.70$ \\ \hline
Skewed-4&$\textbf{0.77}$&$0.41$&$0.38$&$0.39$& $0.64$\\ \hline
\end{tabular}
\end{table}

\begin{table}
\centering
\scriptsize
\caption{\label{tab:comparison-error-rate-image} Comparison of error rate for image recognition}
\begin{tabular}{|l|r|r|r|r|r|} \hline
&adaptivity&JS-divergence&l2-distance&source-accuracy&voting\\\hline \hline
top-1 error&$\textbf{20\%}$&$40\%$&$40\%$&$60\%$& $80\%$\\  \hline
top-2 error&$\textbf{0\%}$&$\textbf{0\%}$&$\textbf{0\%}$&$40\%$& $80\%$\\ \hline
\end{tabular}
\end{table}

\noindent
\textbf{Effectiveness of the Two-Level Index.} We also compare the effectiveness of using the two-level index to the pair-wise strategy for computing the adaptivity metric, as illustrated in Tab.~\ref{tab:image-strategy}. For this scenario, each partition has $200$ samples. The two-level index is observed to achieve more than $8$ times speedup while achieving similar accuracy compared to the pair-wise strategy. The speedup is relatively smaller than the activity recognition scenario. That's mainly because we involve fewer datasets and fewer partitions in this scenario.

\begin{table}
\centering
\scriptsize
\caption{\label{tab:image-strategy} Comparison of adaptivity computation strategies for image recognition}
\begin{tabular}{|l|r|r|r|} \hline
          &top-1 error&top-2 error &latency (milli-sec)\\\hline \hline
pair-wise &$20\%$&$\textbf{0\%}$&$14,585$\\ \hline
2-level LSH w/ flattening&$\textbf{0\%}$&$\textbf{0\%}$&$\textbf{1,652}$\\
\hline
\end{tabular}
\end{table}

\subsubsection{Text Classification}
In this experiment, as mentioned in Sec.~\ref{workloads}, each dataset is pre-processed to remove stop words and then tokenized. For each target dataset, a model will be selected from models that are pre-trained on six source datasets. We compare the Pearson correlation coefficient values between the metrics given by measurements of the \texttt{adaptivity}, the JS-divergence, the L2 distance, source accuracy, and voting strategy on each source dataset, to the target serving accuracy for the corresponding text classification task. We also compare the top-$1$ and top-$2$ error rates.

\begin{table}
\centering
\scriptsize
\caption{\label{tab:comparison-to-text-classification} Comparison of Pearson correlation coefficient (PCC) for text classification (the best Pearson correlation coefficient values are highlighted in bold).}
\begin{tabular}{|l|p{1cm}|p{1cm}|p{1cm}|p{1cm}|p{1cm}|} \hline
target&pcc of adaptivity&pcc of JS-divergence&pcc of l2-distance&pcc of source-accuracy&pcc of voting\\\hline \hline
Financial\_News\_1&$0.18$&$0.49$&$\textbf{0.80}$&$0.21$&$0.39$\\ \hline
Financial\_News\_2&$\textbf{0.93}$&$0.82$&$0.92$&$0.59$&$0.44$\\ \hline
Hate\_Tweets&$0.78$&$0.73$&$\textbf{0.79}$&$0.78$&$0.70$\\ \hline
Tweets\_1&$0.16$&$0.24$&$0.34$&$0.45$&$\textbf{0.55}$\\ \hline
Tweets\_2&$0.34$&$0.26$&$0.23$&$\textbf{0.39}$&$0.01$\\ \hline
IMBD\_Movie\_1&$\textbf{0.29}$&$0.24$&$0.19$&$0.06$&$0.06$\\ \hline
IMBD\_Movie\_2&$\textbf{0.40}$&$0.39$&$0.36$&$0.11$&$0.08$\\ \hline

\end{tabular}
\end{table}

\begin{table} 
\centering
\scriptsize
\caption{\label{tab:comparison-error-rate-text-classification} Comparison of error rate for Text Classification}
\begin{tabular}{|l|r|r|r|r|r|r} \hline
&adaptivity&JS-divergence&l2-distance&source-accuracy&voting\\\hline \hline
top-1 error&$42\%$&$28\%$&$42\%$&$\textbf{14\%}$&$\textbf{14\%}$\\ \hline
top-2 error&$14\%$&$28\%$&$14\%$&$\textbf{0\%}$&$\textbf{0\%}$\\ \hline

\end{tabular}
\end{table}

The results are illustrated in Tab.~\ref{tab:comparison-to-text-classification} and Tab.~\ref{tab:comparison-error-rate-text-classification}, which show that the \texttt{adaptivity} measurement has a higher correlation with the target with most of the candidate models as compared to other metrics, however, source accuracy and voting can more effectively predict the best model for serving with least error rate for all seven search scenarios.

In addition, we observed that the \texttt{adaptivity} metric can be computed within one minute's time in total for all model search scenarios in this workload, while the voting approach, which is the most time-consuming strategy considered in this paper, takes more than ten minutes. Moreover, the JSD-LSH and the L2 distance are significantly faster than the \texttt{adaptivity} measurement, while the source-accuracy approach is the fastest if the source accuracy information has been pre-collected.

\subsubsection{Entity Matching}
In this experiment, different from activity recognition, most attributes contain text-based values. We represent the training dataset of each entity matching task as a bag of words over a shared dictionary for computing the JS-divergence and \texttt{adaptivity}. We compare the Pearson correlation coefficient values of the JS-divergence metric, the \texttt{adaptivity} metric, and source accuracy, to the target accuracy for each EM task, as illustrated in Tab.~\ref{tab:comparison-to-model-adaptivity-em}. We also compare the overall accuracy in terms of top1-error and top2-error for all fifteen tasks, as illustrated in Tab.~\ref{tab:comparison-error-rate-em}. The results show that the \texttt{adaptivity} metric outperforms other metrics in selecting the models to serve with the best accuracy.

\begin{table}
\centering
\scriptsize
\caption{\label{tab:comparison-to-model-adaptivity-em} Comparison of Pearson correlation coefficient of various metrics to the target accuracy for entity matching (the best Pearson correlation coefficient values are highlighted in bold)}
\begin{tabular}{|l|p{1.6cm}|p{1.6cm}|p{1.6cm}|} \hline
target&pcc of adaptivity&pcc of JS-divergence& pcc of source-accuracy\\\hline \hline
Abt\_Buy&$\textbf{0.99}$&$\textbf{0.99}$&$\textbf{0.99}$\\ \hline
Dplp\_Acm&$0.92$&$\textbf{0.94}$&$0.70$\\ \hline
Dblp\_Scholar&$0.91$&$\textbf{0.94}$&$0.83$\\ \hline
Walmart\_Amazon&$\textbf{0.99}$&$\textbf{0.99}$&$0.73$ \\ \hline
MyAnimeList\_AnimePlanet&$\textbf{0.61}$&$0.53$&$0.51$\\ \hline
Bikedekho\_Bikewale&$\textbf{0.51}$&$0.16$&$0.19$\\ \hline
Amazon\_Barnes&$\textbf{0.24}$&$0.17$&$0.22$\\ \hline
GoodReads\_Barnes&$\textbf{0.61}$&$0.53$&$0.20$\\ \hline
Barnes\_Half&$0.29$&$\textbf{0.44}$&$0.31$\\ \hline
RottenTomatoes\_IMDB&$0.11$&$\textbf{0.34}$&$0.16$\\ \hline
IMDB\_TMD&$0.49$&$\textbf{0.67}$&$0.44$\\ \hline
IMDB\_RottenTomatoes&$0.51$&$\textbf{0.75}$&$0.22$\\ \hline
Amazon\_RottenTomatoes&$0.60$&$\textbf{0.64}$&$0.38$\\ \hline
RogerElbert\_IMDB&$\textbf{0.58}$&$0.13$&$0.07$\\ \hline
YellowPages\_Yelp&$\textbf{0.82}$&$0.67$&$0.44$\\ \hline
\end{tabular}
\end{table}

\begin{table}
\centering
\scriptsize
\caption{\label{tab:comparison-error-rate-em} Comparison of error rate for entity matching}
\begin{tabular}{|l|r|r|r|} \hline
&adaptivity&JS-divergence&source-accuracy\\\hline \hline
top-1 error&$\textbf{13\%}$&$47\%$&$33\%$\\ \hline
top-2 error&$\textbf{0\%}$&$27\%$&$27\%$\\ \hline
\end{tabular}
\vspace{-5pt}
\end{table}

\subsubsection{Natural Language Processing}
For this experiment, as mentioned in Sec.~\ref{workloads}, because each task involves merely three datasets, a model discovery scenario for a target dataset only requires comparing two datasets. Therefore, instead of computing the Pearson coefficient for each search scenario, we choose to compute the Pearson coefficient between the metrics and the overall accuracy of each of the two tasks. Each task consists of three different model discovery scenarios. For the same reason, we only consider the top-$1$ error for this experiment, which is counted over all six model discovery scenarios across the two tasks. 
The results are illustrated in Tab.~\ref{tab:comparison-to-model-adaptivity-text} and Tab.~\ref{tab:comparison-error-rate-text}, which show that while \texttt{adaptivity} has less correlation with the target accuracy compared to JS-divergence, it can more effectively predict the best model for serving with zero error rate for all six search scenarios. The source accuracy performs significantly worse than other metrics.

\begin{table}
\centering
\scriptsize
\caption{\label{tab:comparison-to-model-adaptivity-text} Comparison of Pearson correlation coefficient for NLP (the best Pearson correlation coefficient values are highlighted in bold)}
\begin{tabular}{|l|p{1.6cm}|p{1.6cm}|p{1.6cm}|} \hline
target&pcc of adaptivity&pcc of JS-divergence&pcc of source-accuracy\\\hline \hline
Task1&$0.61$&$\textbf{0.71}$&$0.02$\\ \hline
Task2&$0.76$&$\textbf{0.87}$&$0.10$\\ \hline
\end{tabular}
\end{table}

\begin{table}
\centering
\scriptsize
\caption{\label{tab:comparison-error-rate-text} Comparison of error rate for NLP}
\begin{tabular}{|l|r|r|r|} \hline
&adaptivity&JS-divergence&source-accuracy\\\hline \hline
top-1 error&$\textbf{0\%}$&$16.7\%$&$33.3\%$\\ \hline
\end{tabular}
\end{table}

\subsection{The Impact of Hyperparameter Tuning}

We further evaluate how hyperparameters, such as $r$ (as in Eq.~\ref{eq:eq1}), the number of concatenated hash functions (i.e., $K$) of each band, the number of bands (i.e., $L$), and the number of hash bins for creating probability distribution for comparison, will all affect the accuracy of the JS-divergence computed using LSH compared to the baseline, which are illustrated in Fig.~\ref{fig:ar-tuning}. It shows that it is possible to find a set of parameters that ensure high precision and recall by tuning the hyperparameters associated with the LSH scheme. %

\subsection{The Impact of Partitioning}

\begin{figure}
\centering\subfigure[JS-divergence]{%
   \label{fig:jsd-latency}
   \includegraphics[width=1.58in]{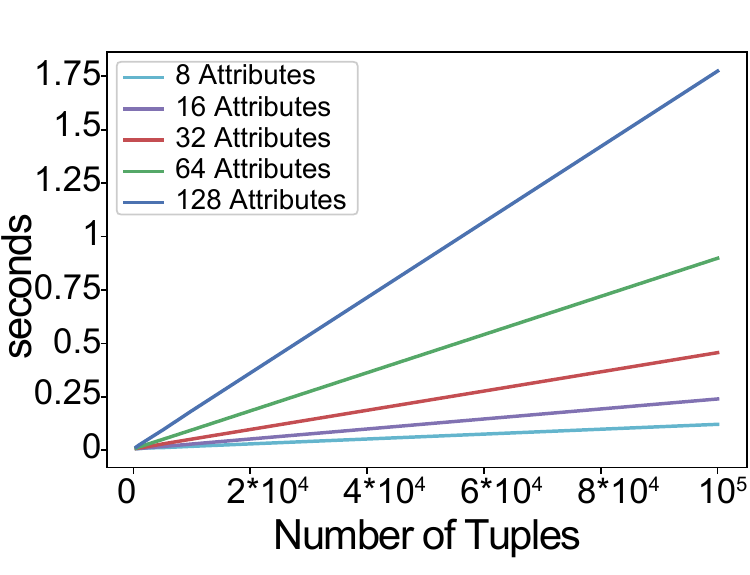}  
}%
\hspace{0pt}
\subfigure[adaptivity]{%
  \label{fig:adaptivity-latency}
  \includegraphics[width=1.58in]{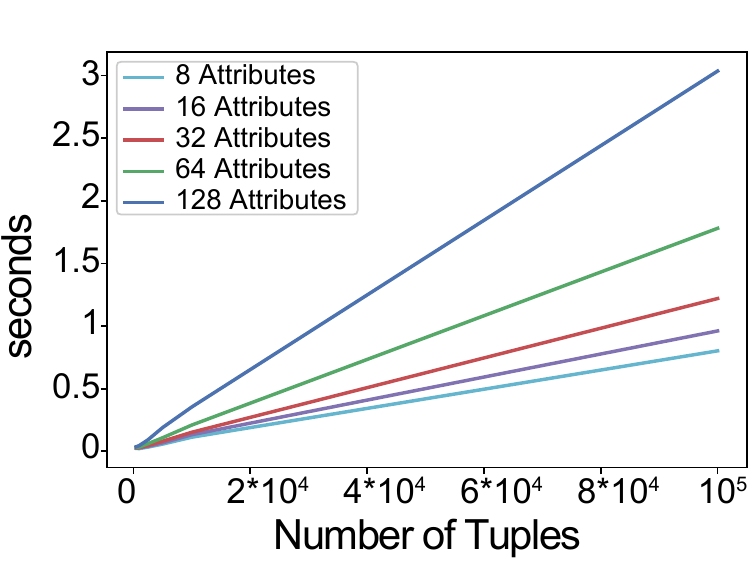}
}%

\vspace{-3pt}
\subfigure[2500 rows \& 32 cols for source/target]{%
  \label{fig:adaptivity-latency}
  \includegraphics[width=1.58in]{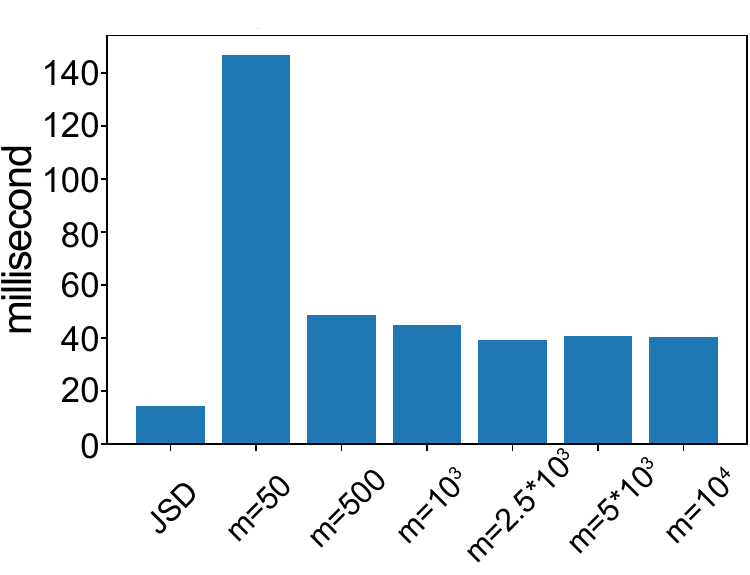}
}%
\hspace{0pt}
\subfigure[10000 rows \& 32 cols for source/target]{%
  \label{fig:adaptivity-latency}
  \includegraphics[width=1.58in]{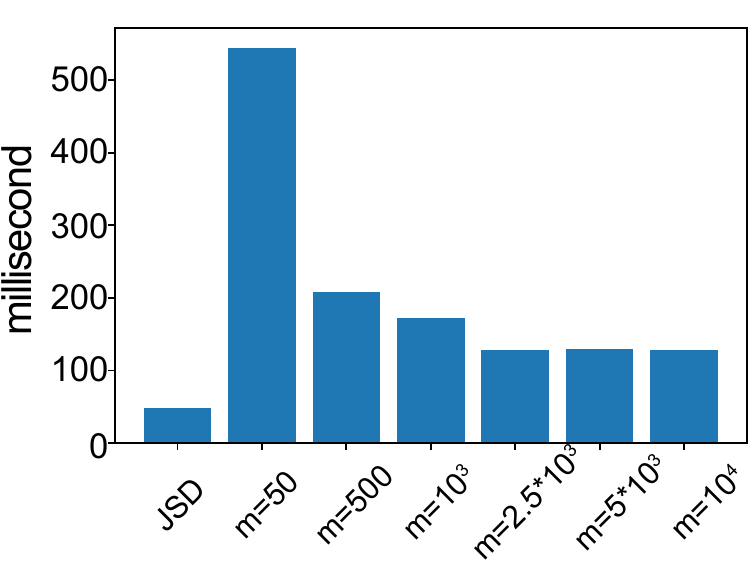}
}%
\eat{
\vspace{1pt}
\subfigure[latency comparison]{%
  \label{fig:adaptivity-latency}
  \includegraphics[width=1.58in]{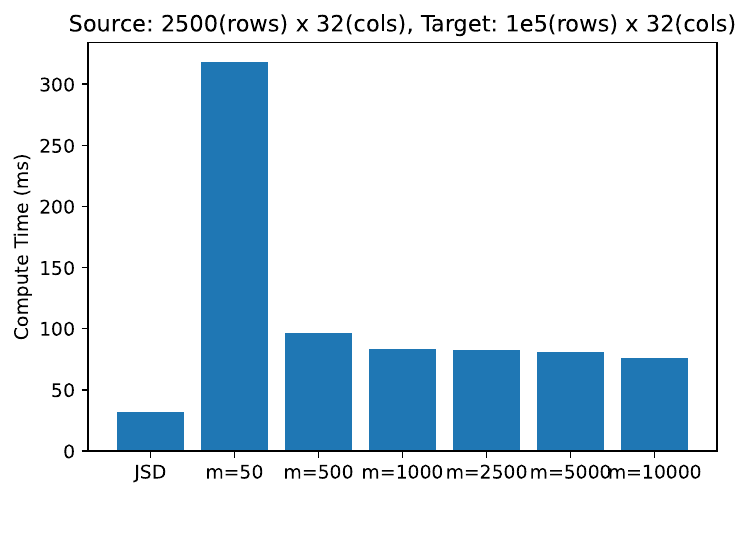}
}%
\hspace{0pt}
\subfigure[latency comparison]{%
  \label{fig:adaptivity-latency}
  \includegraphics[width=1.58in]{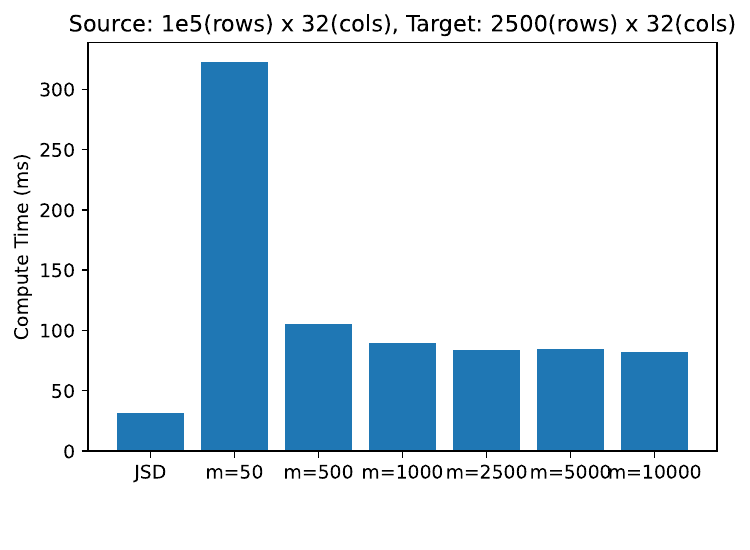}
}%
}

\caption{\label{fig:latency}
Latency comparison of computing JS-divergence and adaptivity, m is the size of each partition
}
\end{figure}

One thing to note is that the computation of \texttt{adaptivity} is more complicated than JS-divergence and takes more time. For most of the above experiments, we choose the partition size of each dataset between $300$ to $800$. For the NLP experiment, when the source dataset and the target dataset have a significant size discrepancy, we choose the size of the smaller dataset to be the partition size, so that the smaller dataset has only one partition. Then we compare the latency of computing JS-divergence and \texttt{adaptivity} by using different partition sizes and for different sizes of the source and target datasets. The results are illustrated in Fig.~\ref{fig:latency}, which illustrates that when partition size is around $500$, the computational overhead for \texttt{adaptivity} is $1.7-3\times$ of JS-divergence. However, the latency of computing \texttt{adaptivity} is still $10\times$ lower than the runtime latency of the voting approach which requires each candidate model to run prediction on each sample in the target dataset.

\begin{figure}
\centering
{\subfigure[r]{%
   \label{fig:r}
   \includegraphics[width=3.3in]{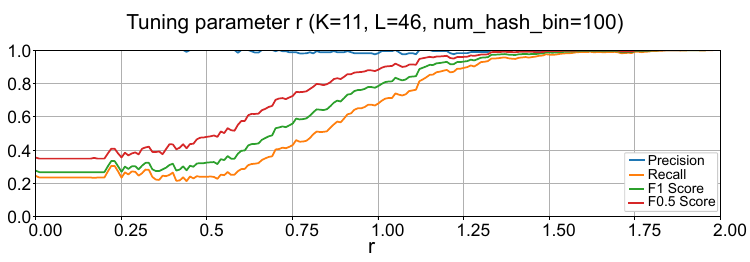}  
}%
\vspace{-1pt}
\subfigure[number of concatenated hash functions]{%
  \label{fig:K}
  \includegraphics[width=3.3in]{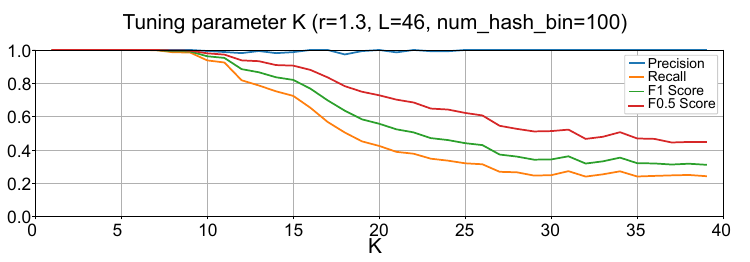}
}%
\vspace{-1pt}
\subfigure[number of hash tables]{%
  \label{fig:L}
  \includegraphics[width=3.3in]{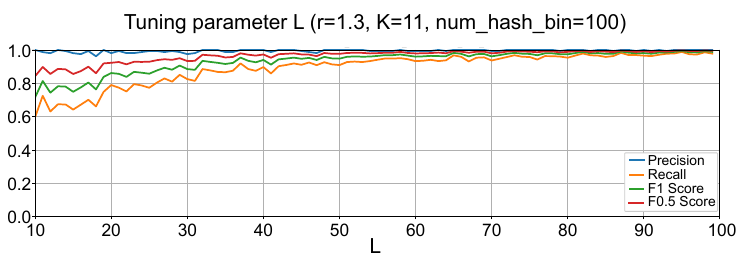}
}%
\vspace{-1pt}
\subfigure[number of hash bins]{%
  \label{fig:L}
  \includegraphics[width=3.3in]{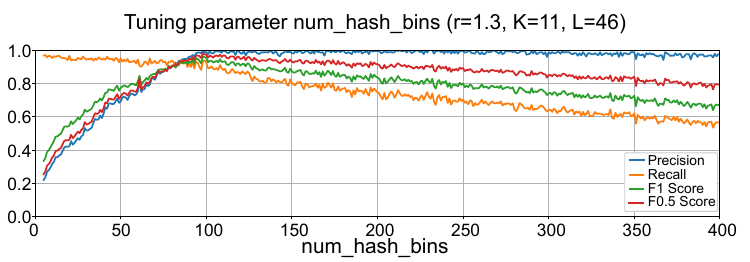}
}
}
\caption{\label{fig:ar-tuning}
Hyperparameter tuning for LSH of JS-divergence.
}

\end{figure}

%% file: relatedworks.tex
\section{Related Works}
AutoML methods, such as Auto-WEKA~\cite{thornton2013auto}, Auto-sklearn~\cite{feurer2020auto}, TPOT~\cite{olson2016tpot}, H2O~\cite{ledell2020h2o} automatically search ML model algorithms as well as hyper-parameters. Although their works can also be applied to our targeting problem by modeling the model search process as an optimization problem that minimizes the loss functions defined over the targeting domain, they are mainly designed for the training stage, and their searching strategies require labels of target data for evaluating the loss during the optimization process. In this work, we mainly focus on the deployment stage where it is challenging to collect and manage sufficient labeled data for AutoML tasks.

Clipper~\cite{crankshaw2017clipper} proposes a strategy to select models for ensemble inference by modeling the scenario as a multi-armed bandit problem. DJEnsemble~\cite{pereira2021djensemble} presents a cost-based approach for the automatic selection of black-box models to answer spatio-temporal queries. We do not consider such approaches in this work, mainly because they all require labeled data from the target domain for model selection, which could become a significant burden that complicates the model deployment phase.

In recent, numerous works are proposed to address open data discovery problems, including automatically discover table unionability~\cite{nargesian2018table} and joinability~\cite{zhu2016lsh,zhu2019josie}, and related tables~\cite{zhang2020finding}. Most of these works are trying to identify all similar features using LSH techniques based on Jaccard similarity or variants. While these works can be leveraged to accelerate the match of JSD-LSH signatures, they are not directly applicable in selecting related models, because the existence of a significant portion of shared features between the source and target datasets doesn't mean the probability distributions over the two feature spaces are similar. 

Zamir and et al.~\cite{zamir2018taskonomy} propose a computational approach to find the transferable relationships, abstracted as taxonomy among the different computation vision tasks, e.g., depth estimation, edge detection, point matching and etc., so that some tasks can be trained using other tasks' output and thus require less training data and supervision budget. Their work is focused on the adaptivity from a task's output domain to another task's input domain. In contrast, our work is focused on the adaptivity of tasks' input domains. Also, while their work is limited to computer vision, our work is targeting a more generalized scenario. %